\documentclass[prb,twocolumn,showpacs,floatfix,titlepage,superscriptaddress]{revtex4-1}

\usepackage{graphicx}
\usepackage{amsmath}
\usepackage{amsfonts}
\usepackage{hyperref}
\hypersetup{colorlinks, linkcolor={blue},citecolor={blue}, urlcolor={blue}}

\begin{document}

\title{Tight-binding couplings in microwave artificial graphene}

\author{Matthieu Bellec}
\affiliation{Universit\'{e} Nice Sophia Antipolis, CNRS, Laboratoire de Physique de la Mati\`ere Condens\'ee, UMR 7336, 06100 Nice, France}
\author{Ulrich Kuhl}
\affiliation{Universit\'{e} Nice Sophia Antipolis, CNRS, Laboratoire de Physique de la Mati\`ere Condens\'ee, UMR 7336, 06100 Nice, France}
\author{Gilles Montambaux}
\affiliation{Universit\'{e} Paris-Sud, CNRS, Laboratoire de Physique des Solides, UMR 8502, 91405 Orsay Cedex, France}
\author{Fabrice Mortessagne}
\email{fabrice.mortessagne@unice.fr}
\affiliation{Universit\'{e} Nice Sophia Antipolis, CNRS, Laboratoire de Physique de la Mati\`ere Condens\'ee, UMR 7336, 06100 Nice, France}

\date{\today}

\begin{abstract}
We experimentally study the propagation of microwaves in an artificial honeycomb lattice made of dielectric resonators. This evanescent propagation is well described by a tight-binding model, very much like the propagation of electrons in graphene. We measure the density of states, as well as the wave function associated with each eigenfrequency.  By changing the distance between the resonators, it is possible to modulate the amplitude of next-(next-)nearest-neighbor hopping parameters and to study their effect on the density of states.  The main effect is the density of states becoming dissymmetric  and a shift of the energy of the Dirac points. We study the basic elements: An isolated resonator, a two-level system, and a square lattice. Our observations are in good agreement with analytical solutions for corresponding infinite lattice.
\end{abstract}
\pacs{42.70.Qs, 03.65.Nk, 71.20.-b, 73.22.Pr}

\maketitle

\section{Introduction}
Artificial graphene\cite{Polini2013} is an emerging field which offers a playground to investigate physical phenomena related to massless Dirac fermions in situations hardly reachable in genuine graphene. As reported recently,\cite{Polini2013} many different low-energy physical systems such as 2D electron gas,\cite{Singha2011} ultracold atoms in optical lattice,\cite{Wunsch2008,*Montambaux2009,*Soltan-Panahi2011,Tarruell2012,*Lim2012} molecular assembly,\cite{Gomes2012} and photonic crystals constitute pertinent candidates.\cite{Peleg2007,Sepkhanov2007,*Haldane2008,Kuhl2010,Bittner2010,*Bittner2012,Bellec2013,Rechtsman2013b,Rechtsman2013,Khanikaev2013} In such artificial systems, the periodicity of the lattice induces an energy band structure very similar to the one encountered in condensed-matter crystals. When two sites per unit cell and a triangular symmetry are considered -- i.e.,~a honeycomb lattice (hc) -- conical singularities, the so-called Dirac points, may emerge at the corner of the first Brillouin zone in an analogous manner to what happens in the electronic spectrum of graphene.\cite{CastroNeto2009} The key advantage of these systems resides in the high flexibility and control regarding the lattice properties. Consequently, numerous phenomena have been recently observed ranging from edge-state observation~\cite{Kuhl2010,Rechtsman2013} in regular lattices to topological phase transition of Dirac points~\cite{Tarruell2012,Bellec2013,Rechtsman2012b} and Landau level creation~\cite{Gomes2012,Rechtsman2013b} in strained lattices.

\begin{figure}
\centering
\includegraphics[width=8.6cm]{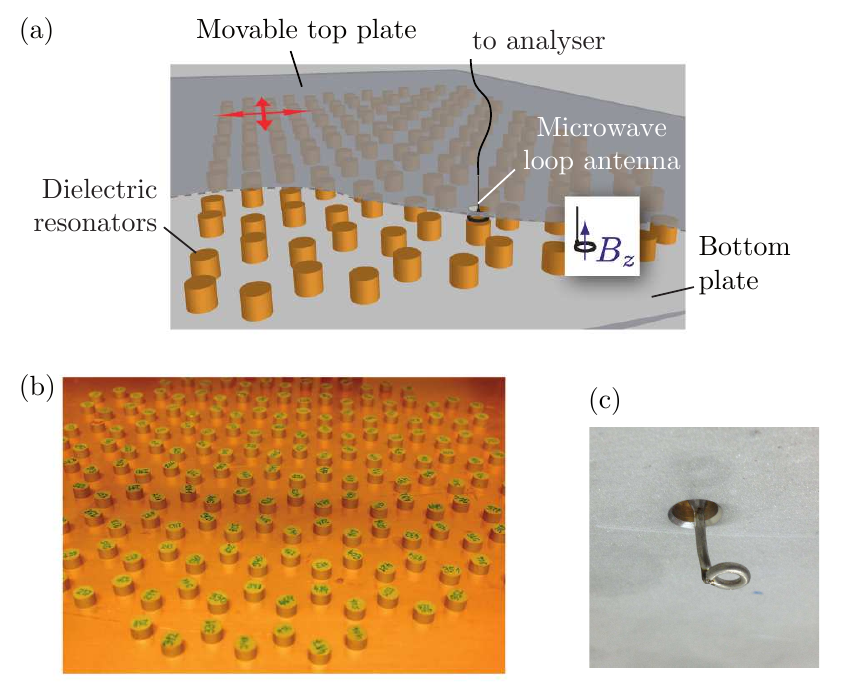}
\caption{\label{fig:ExpSetup}(Color online) Typical experimental setup used to realize a tight-binding microwave analog. (a) Sketch of the setup. A dielectric lattice structure is inserted in between two metallic plates. A loop antenna crossing the top plate (inset) and connected to a vectorial network analyzer is used to generate and collect the microwave signal. A scanning system allows to move the top plate. (b) Picture of the dielectric structure (top plate removed). (c) Picture of the loop antenna.}
\end{figure}

Most of the observations are usually modeled with tight-binding (TB) theory\cite{Wallace1947,Reich2002} and include only the nearest-neighbor (N1) coupling terms which are the most dominant ones. While generally ignored, next-to-nearest-neighbor (N2) and third-nearest-neighbor (N3) coupling terms are not negligible in graphene.\cite{CastroNeto2009} For instance, the ratio between N1 and N2 coupling is of the order of 5\% and can be even larger in bilayer or doped graphene.\cite{Reich2002,McChesney2010} Higher-order couplings can shift the Dirac points or generate dissymmetric band structures\cite{Reich2002,McChesney2010,CastroNeto2009,Bena2011} and modify the properties of the edge states in both monolayer ribbons\cite{Sasaki2006} and bilayer graphene.\cite{Cortijo2010} Recent works have proposed to play on the N3/N1 coupling ratio to create and move Dirac points.\cite{Bena2011,Hasegawa2012,Montambaux2012,Sticlet2013}

In this paper, we use a photonic artificial graphene, working in the microwave range, to experimentally probe the role of high-order coupling terms in the frame of the TB regime (see Fig.~\ref{fig:ExpSetup}). The N1, N2, and N3 coupling terms can be varied by changing the lattice constant. When increasing the coupling terms beyond nearest neighbors, we observe a modification of the density of states (DOS): The spectrum becomes dissymmetric and  the energy of the Dirac point is shifted. However, the salient features of the DOS -- two bands, a vanishing (Dirac) point and two logarithmic divergences -- remain unchanged.

The paper is organized as follows. To well establish the tight-binding regime, we first describe, in Sec.~\ref{sec:MicrowaveTB}, the response of the basic elements: (i) an isolated resonator and (ii) two weakly coupled resonators. Two lattices, square and honeycomb, composed of a few hundreds of identical resonators are then considered. In both cases we present the DOS and the associated eigenstates obtained through local density of states (LDOS) measurements. In Sec.~\ref{sec:NextNN}, we emphasize the importance of the higher-order nearest-neighbors coupling terms. We discuss how these parameters affect the DOS by comparing experimental spectra and analytical calculations for infinite structures. We draw a conclusion in Sec.~\ref{sec:Conclusion}.

\section{Microwave nearest-neighbors tight-binding analog}
\label{sec:MicrowaveTB}
\subsection{Experimental setup}
Figure~\ref{fig:ExpSetup}(a) presents a sketch of the typical experimental setup.\cite{Kuhl2010,Barkhofen2013} Two metallic plates, separated by 17\,mm, constitute the electromagnetic (EM) cavity. A set of identical cylindrical  resonators is placed in between. Each resonator has a radius $r_D$ = 4\,mm, a height of 5\,mm, and a high permittivity $\epsilon$ = 36 (i.e.,~refractive  index $n=6$). Figure~\ref{fig:ExpSetup}(b) shows a picture of such a structure (note that the top plate has been removed). A single-loop antenna [see Fig.~\ref{fig:ExpSetup}(c)] goes through the top plate. The geometry of the system allows to excite only the lowest TE mode inside the resonators. Typically, the cut-off frequencies are about 5 GHz inside the dielectric resonator and 10 GHz outside. The evanescent field in the air  ensures a weak-coupling regime between resonators.\cite{Kuhl2010} The different couplings will be carefully analyzed in the following two sections. The microwave signal is generated and collected using a standard vectorial network analyzer providing the scattering matrix $S$. The measured quantity is, in our case, the reflected signal $S_{11}$. Note that the bottom plate is fixed while the top plate is movable. Thus, it is worth mentioning that, compared to our previous experimental setup,\cite{Kuhl2010,Barkhofen2013} this configuration allows for a full scan of the EM field all over the structure. As we will see, this setup allows us to have access to both the DOS (i.e.,~eigenfrequencies) and the associated eigenstates.

\subsection{The basic element: An isolated resonator}

\begin{figure}[t]
\centering
\includegraphics[width=8.6cm]{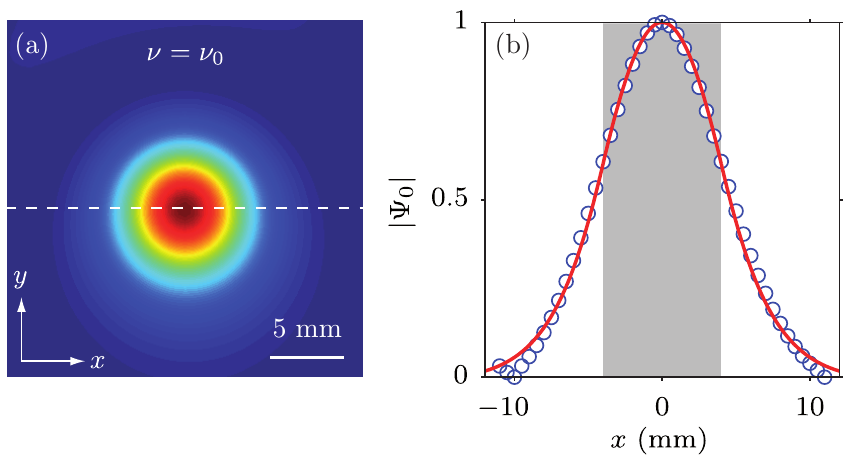}
\caption{\label{fig:OneDisc} (Color online) Isolated resonator response. (a) Normalized experimental wave function intensity $|\Psi_0|^2$ at the frequency $\nu_0 = 6.65$\,GHz. (b) Profile of $|\Psi_0|$ corresponding to the dashed line in (a). The gray zone is the disk position. Red curve: Fit from Eq.~(\ref{eq:Bz2}) with $\gamma_j = 0.3341$\,mm$^{-1}$, $\gamma_{k,1} = 0.1215$\,mm$^{-1}$, $\gamma_{k,2} = 0.3423$\,mm$^{-1}$, and $\gamma_{k,3} = 0.5366$\,mm$^{-1}$ and $\alpha_1 = -0.1370$, $\alpha_2 = 5.3816$, and $\alpha_3 = -6.6585$.}
\end{figure}

Due to Mie resonance,\cite{Lidorikis1998,*Laurent2007a} the reflected signal of an isolated resonator exhibits a peak centered at $\nu_0 = 6.65$\,GHz. In the ideal case, where the spacing $h$ between the two metallic plates corresponds to the height of resonators, the system has a cylindrical symmetry, thus allowing to separate the $z$ and radial coordinates. Here we concentrate on the TE mode, where the wave function $\Psi_0$ corresponds to the $z$ component of the magnetic field at $\nu_0$ as:\cite{Kuhl2010} $B_z(r,z) = B_0 \sin \left(\frac{\pi}{h} z\right) \Psi_0(r)$ with
\begin{equation}
\label{eq:Bz}
\Psi_0(r) =
\begin{cases}
 J_0(\gamma_j r) & \text{if $r < r_D$}, \\
 \alpha K_0(\gamma_k r) & \text{if $r > r_D$},
\end{cases}
\end{equation}
where $\Psi_0(0) = 1$. $J_0$ and $K_0$ are Bessel functions, $r$ is the distance from the center of the disk, $\gamma_{j} = \sqrt{\left(\frac{2 \pi \nu_0 n}{c}\right)^2 - \left(\frac{\pi}{h}\right)^2}$, and $\gamma_{k} = \sqrt{\left(\frac{\pi}{h}\right)^2-\left(\frac{2 \pi \nu_0}{c}\right)^2}$ ($n$ denoting the refractive index).
In our case $h = 17\,$\,mm, which means that the upper plate has a non-negligible distance to the disk, so that the cylindrical symmetry is lost. Due to the three-dimensionality, the field inside the disk can excite several evanescent TE modes outside. Their corresponding wavenumber is given by $\gamma_{k,m} = \sqrt{\left(\frac{m \pi}{h}\right)^2-\left(\frac{2 \pi \nu_0}{c}\right)^2}$. Finally, we assume the following:
\begin{eqnarray}
\label{eq:Bz2}
B_z(r,z) & \approx & B_0 \Psi_0(r,z) \nonumber \\
&=&
\begin{cases}
f(z)  J_0(\gamma_j r) & \text{if $r < r_D$}, \\
\displaystyle \sum_m \alpha'_m \sin \left(\frac{m \pi}{h} z\right) K_0(\gamma_{k,m} r) & \text{if $r > r_D$}.
\end{cases}
\end{eqnarray}
$f(z)$ describes the $z$ dependence of the magnetic field and verifies the boundary conditions $f(0) = f(h) = 0$. It takes into account the fact that $h$ is larger than the disk height (5\,mm). $\gamma_{j}$ is now defined via the function $f$. $\gamma_{k,m}$ is calculated using $h$ and the measured eigenfrequency of the disk $\nu_0$. The loop antenna is sitting at a fixed height $z_0$ and for simplicity we include the $z$-dependence and the normalization in $\alpha_{m}=\alpha'_m \sin \left(\frac{m \pi}{h} z_0\right)/f(z_0)$. The coefficients are obtained by a fitting procedure including continuity conditions.

As detailed in the Appendix~\ref{app:gfunction}, $|\Psi_0(\textbf{r}_1)|$ is related to the reflection signal $S_{11}(\nu)$ ($\textbf{r}_1$ denoting the position of the antenna) through a Breit-Wigner function, at the vicinity of the resonance $\nu_0$, as follows:
\begin{equation}
S_{11}(\nu) = 1 - \textrm{i}\sigma \frac{|\Psi_0(\textbf{r}_1)|^2}{\nu - \nu_0 + \textrm{i} \Gamma}
\label{eq:BWfunction}
\end{equation}
where $\sigma$ is a coupling term slowly varying with the frequency and $\Gamma$ corresponds to the spectral width of the resonance essentially due to Ohmic losses ($\sigma \ll \Gamma$). Therefore, by fitting the resonance with a Lorentzian shape, one has access to the wave function $|\Psi_0|$ up to a factor $\sqrt{\sigma}$. Figure~\ref{fig:OneDisc}(a) shows the intensity $|\Psi_0(\textbf{r})|^2$ where $\Psi_0$ is normalized such that $\Psi_0(0)=1$. Figure~\ref{fig:OneDisc}(b) corresponds to the profile $|\Psi_0|$ measured along the $x$ axis [dashed line in Fig.~\ref{fig:OneDisc}(a)]. We observe that the energy is mostly confined within the disk (delimited by the gray zone) and spreads out evanescently. The fit obtained using three evanescent modes is shown as a red solid line in Fig.~\ref{fig:OneDisc}(b). The fit parameters are indicated in the figure caption. Have in mind that the loop antenna is not a point-like antenna. It is integrating over a small surface therefore leading to effective parameters $\gamma_j$ and $\alpha_m$.

\subsection{Two-disk system}
\label{subsec:TwoDisc}

\begin{figure}
\centering
\includegraphics[width=8.6cm]{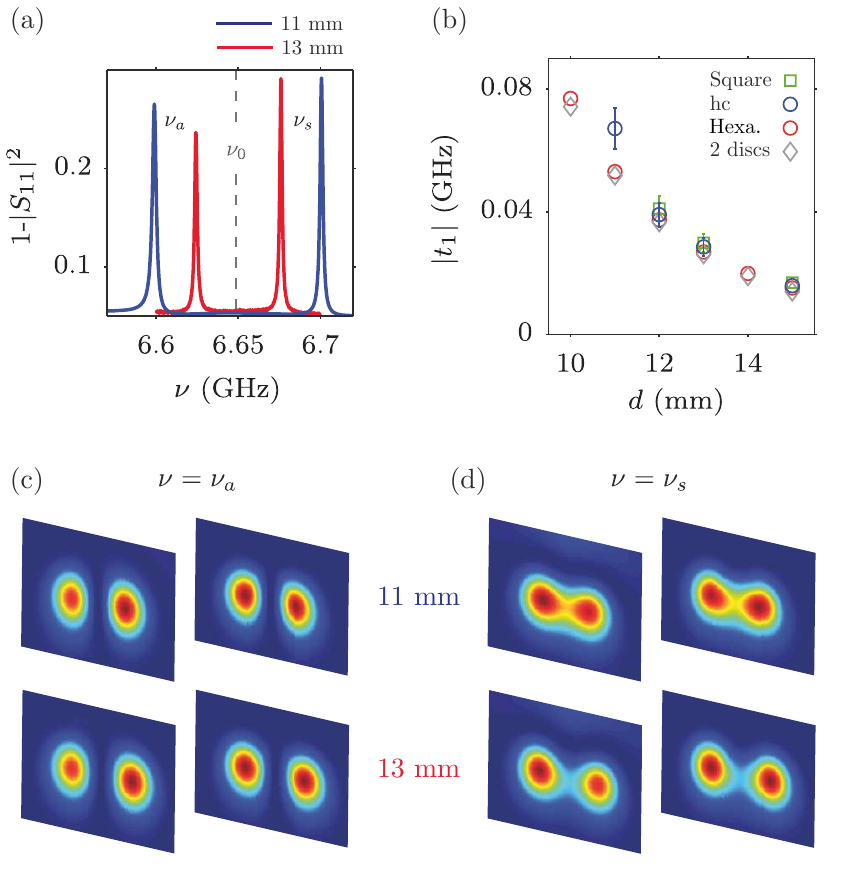}
\caption{\label{fig:TwoDiscs}(Color online). (a) Two-disks response. Frequency splitting, $\Delta\nu = \nu_s - \nu_a$, for $d=11$\,mm and 13\,mm (resp. blue and red curves). The dashed line corresponds to the isolated resonance $\nu_0$. (b) Coupling strength obtained from the experiments (see text for details). If not represented, the error bars are smaller than the symbol size. (c) and (d). Eigenfunction intensity $|\Psi(r)|^2$ at $\nu = \nu_a$ (anti-symmetric) (c) and $\nu = \nu_s$ (symmetric) (d). Left panels in (c) and (d): Experimental wave functions for $d=11$\,mm (top) and 13\,mm (bottom). Right panels in (c) and (d): Linear superposition of isolated wave functions spaced by $d=11$\,mm (top) and 13\,mm (bottom).}
\end{figure}

When two identical resonators are close to each other by a distance less than a few diameters, the evanescent nature of the excited mode outside the dielectric medium leads to a coupling illustrated by a symmetric frequency splitting: $\nu_a=\nu_0-\Delta\nu/2$,  $\nu_s=\nu_0+\Delta\nu/2$ [see Fig.~\ref{fig:TwoDiscs}(a)]. This splitting is nothing else than twice the N1 coupling strength $|t_1|$ and depends on the separation $d$. Therefore, the systematic measurement of $\Delta \nu$ for various $d$ allows obtaining $|t_1(d)|$ [Fig.~\ref{fig:TwoDiscs}(a) actually presents two cases for $d=11$\,mm and 13\,mm].\cite{Kuhl2010,Barkhofen2013} The gray diamonds in Fig.~\ref{fig:TwoDiscs}(b) show extracted $|t_1|$ for few more $d$. It is worth noting that the couplings obtained in a benzene-like system (i.e.,~six disks with hexagonal arrangement) are similar (red circles). As described in Sec.~\ref{sec:NextNN}, the values obtained with the square and the hc lattices (green squares and blue circles, respectively) are also consistent.
For both frequencies $\nu_a$ and $\nu_s$ [resp.\ Figs.~\ref{fig:TwoDiscs}(c) and \ref{fig:TwoDiscs}(d), left panels] and for $d$ = 11\,mm and 13\,mm (resp. top and bottom panels), we measure the wave function intensity $|\Psi(\textbf{r}_i)|^2$ (left panels). It is noticeable that in this experimental setup, the state with lowest frequency corresponds to an antisymmetric configuration for the magnetic field [Fig.~\ref{fig:TwoDiscs}(c)]. Meanwhile the electric field configuration is symmetric.
The two-disk system can be viewed as two weakly coupled isolated resonators, as illustrated in the right panels of Figs.~\ref{fig:TwoDiscs}(c) and \ref{fig:TwoDiscs}(d) where are plotted the difference (anti-symmetric) and the sum (symmetric) of two identical isolated eigenfunctions [Fig.~\ref{fig:OneDisc}(b)] associated with each resonator.

\subsection{Local density of states and wave functions}
\subsubsection*{Square lattice}

\begin{figure*}
\centering
\includegraphics[width=17.8cm]{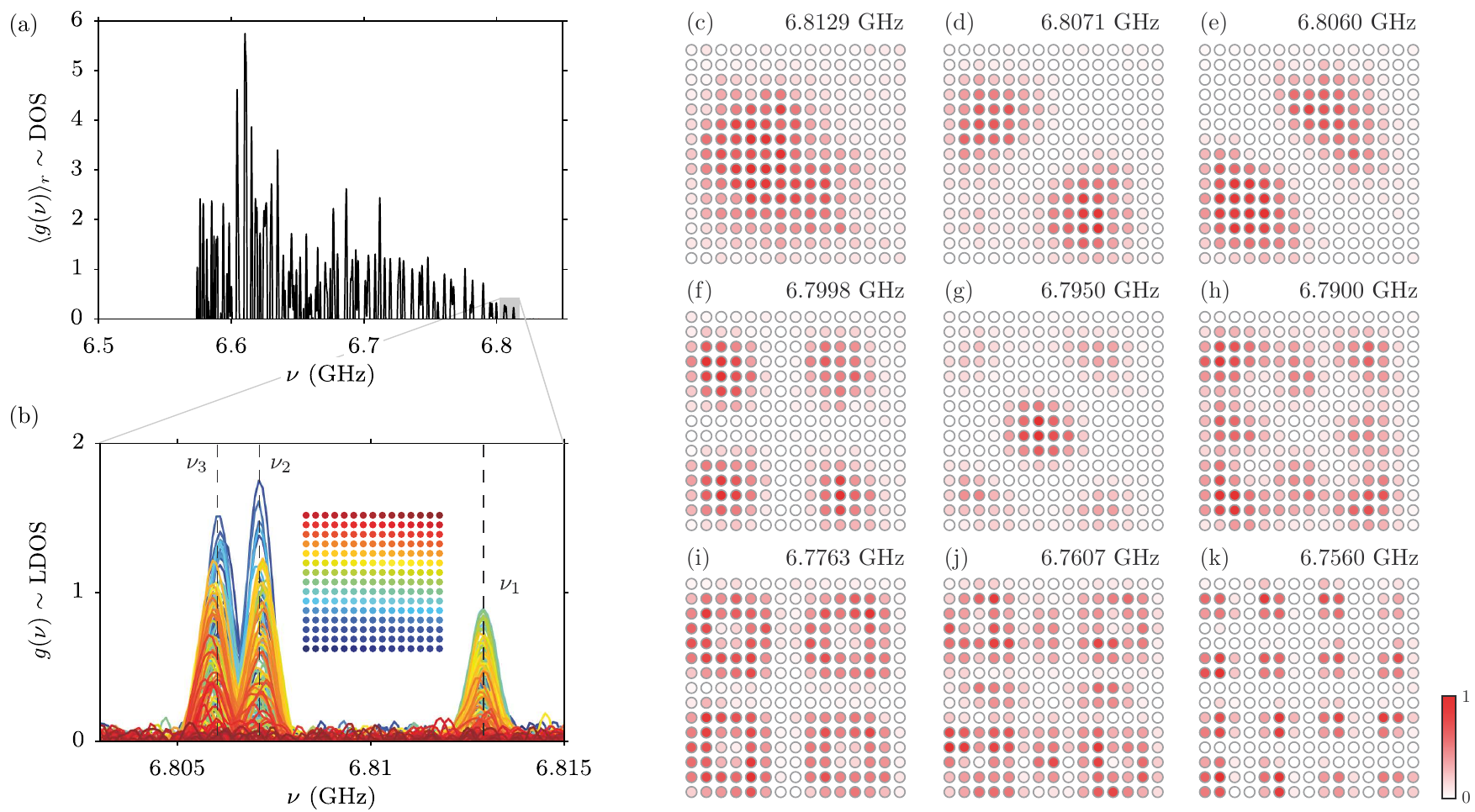}
\caption{\label{fig:LDOSsquare} (Color online) DOS, LDOS, and wave functions for the square lattice with $d$ = 13\,mm spacing. (a) Measured DOS, through $g$ function; see Appendix~\ref{app:gfunction}. (b) Measured LDOSs in a small frequency range corresponding to the gray zone in (a). Each site is marked with a color ranging from deep blue to red (inset). (c)--(k) Experimental wave function intensities for various eigenfrequencies ranging from 6.7560 GHz (k) to 6.8129 GHz (c). (c), (d), and (e) correspond to $\nu_1$, $\nu_2$, and $\nu_3$ in (b), respectively. (d) and (e) are nearly degenerated states (see text for details).}
\end{figure*}

\begin{figure}
\centering
\includegraphics[width=8.6cm]{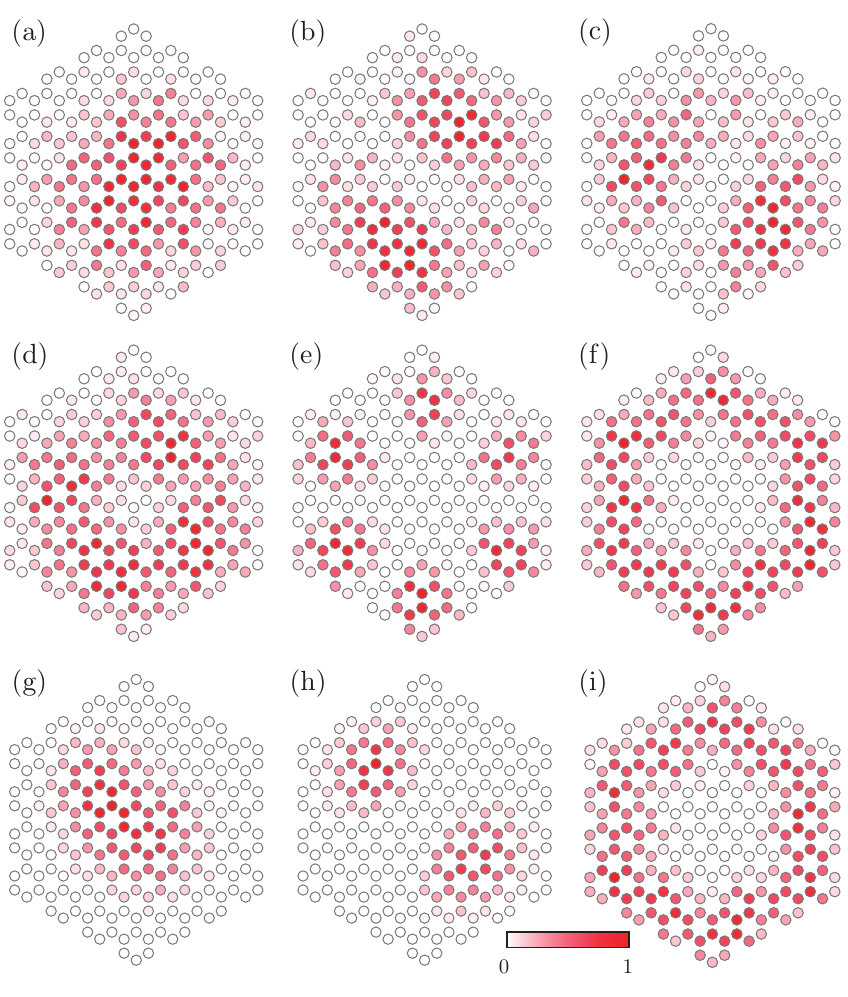}
\caption{\label{fig:LDOS} Experimental wave function intensities for the hc lattice corresponding to different eigenfrequencies. (a)--(h) Lattice constant $d$ = 12\,mm. (a) $\nu = 6.8086$\,GHz. (b)--(c) Nearly degenerated states at $\nu = 6.8037$\,GHz and $\nu = 6.8024$ GHz respectively. The symmetry of the mode is broken. (d) Superposition of the two nearly degenerated states. The symmetry is restored. (e) $\nu = 6.7891$ GHz. (f) $\nu = 6.7789$. (g) $\nu = 6.5611$. (h) $\nu = 6.5622$. (i) Lattice constant $d$ = 11\,mm. $\nu = 6.8334$.}
\end{figure}

Knowing the basic element characteristics, we can now consider larger structures and build a lattice (here with 225 resonators). In this subsection, we first give the experimental details to obtain the LDOS (i.e.,~eigenvalues for each site positions $\mathbf{r}$) and the associated wave functions (i.e.,~eigenstates) in the case of a square lattice with disk separation $d$ = 13\,mm. As discussed in the Appendix~\ref{app:gfunction}, we work with a quantity directly related to the LDOS, namely the $g$ function. Figure~\ref{fig:LDOSsquare}(a) shows the measured DOS obtained by averaging the $g$-function over all the site positions. Details of the LDOS are depicted in Fig.~\ref{fig:LDOSsquare}(b). Each color (from deep blue to red) corresponds to a site position (resp. bottom to top, see the inset). At given eigenfrequencies (e.g. $\nu_1$, $\nu_2$ and $\nu_3$), the LDOS magnitudes associated with each position $\mathbf{r}$ (i.e.,~ with each color) can be picked up. The visualization of the wave function distribution associated with each eigenfrequency thus becomes accessible. Figures~\ref{fig:LDOSsquare}(c)--\ref{fig:LDOSsquare}(k) display the experimental wave function intensities corresponding to various eigenfrequencies ranging from 6.7560 GHz (k) to 6.8129 GHz (c). Figures~\ref{fig:LDOSsquare}(c), \ref{fig:LDOSsquare}(d), and \ref{fig:LDOSsquare}(e) correspond to $\nu_1$, $\nu_2$ and $\nu_3$ in Fig.~\ref{fig:LDOSsquare}(b), respectively. For $\nu_1$, the mode is mostly confined within the bulk with an homogeneous distribution. The global square symmetry is broken for the eigenstates corresponding to $\nu_2$ and $\nu_3$. However, when the two mode intensities are superposed, the symmetry is restored. Such an observation indicates that these eigenstates are nearly degenerated ($|\nu_3-\nu_2|=1.1$ MHz is much smaller than the resonance width $\sim 10$ MHz) and should be degenerated if the square symmetry was perfect. The degeneracy is lifted by the disorder in the bare frequency $\nu_0$, i.e., the eigenfrequency of each resonator, which is distributed within a range of 10 MHz around 6.65 GHz. Figures~\ref{fig:LDOSsquare}(f)--\ref{fig:LDOSsquare}(k) depict the eigenstates at lower eigenfrequencies. Although not presented here, the numerical simulations, performed by diagonalizing the TB Hamiltonian with an appropriate coupling strength, are in very good accordance with the experiments. Note that a slight dissymmetry can be observed in the experimental eigenstates (the modes seem to be shifted to the bottom-left corner). Systematic measurements allow us to attribute this behavior to the anisotropic response of the antenna [essentially due to its straight part perpendicular to the loop; see Fig.~\ref{fig:ExpSetup}(c)]. At this step, we can thus claim that our setup allows an accurate reconstitution of the tight-binding model providing both LDOS and eigenstates. The DOS is simply obtained by averaging the LDOS over all the positions and will be considered in the next section.

\subsubsection*{Honeycomb lattice}
As presented in Fig.~\ref{fig:LDOS}, we perform similar measurements in the case of a hc lattice. Figures~\ref{fig:LDOS}(a)--\ref{fig:LDOS}(h) correspond to a lattice constant $d$ = 12\,mm. Here again, the first mode [Fig.~\ref{fig:LDOS}(a)] is confined within the bulk; the two following modes [Figs.~\ref{fig:LDOS}(b) and \ref{fig:LDOS}(c)] are  (nearly) degenerated. Figure~\ref{fig:LDOS}(d) shows that the global six-fold symmetry is restored when these two modes are superposed. Moreover, we observe that the two highest frequency modes [Figs.~\ref{fig:LDOS}(a) and \ref{fig:LDOS}(c)] are very similar to the two lowest frequency ones [Figs.~\ref{fig:LDOS}(g) and \ref{fig:LDOS}(h)]. This behavior will be commented on in the next section. Finally, Figs.~\ref{fig:LDOS}(f) and \ref{fig:LDOS}(i) show an example of two modes with the same index in the respective spectra of hc lattices with two different lattice constants ($d$= 11 and 12\,mm). The change of the coupling strength leads to a shift of the eigenfrequencies however; it does not affect the eigenstates.

\section{Higher-order nearest-neighbor couplings}
\label{sec:NextNN}
From the spectra presented in Fig.~\ref{fig:LDOSsquare}(a), one can extract another crucial piece of information: The density of states. As shown in Appendix~\ref{app:gfunction}, the $g$ function averaged over all the site positions is a quantity directly related to the DOS. If we restrict the TB model to the N1 interactions, we expect to have a symmetric DOS in both square and hc lattices. This is opposed to what we observe in Fig.~\ref{fig:LDOSsquare}(a), where the LDOSs are clearly not symmetric. In this section, we will emphasize the role of higher order nearest-neighbor coupling terms and show, both experimentally and analytically, how significant they are in the DOS shape modification.

\subsection{Tight-binding Hamiltonian}

\begin{figure}[t]
\centering
\includegraphics[width=8.6cm]{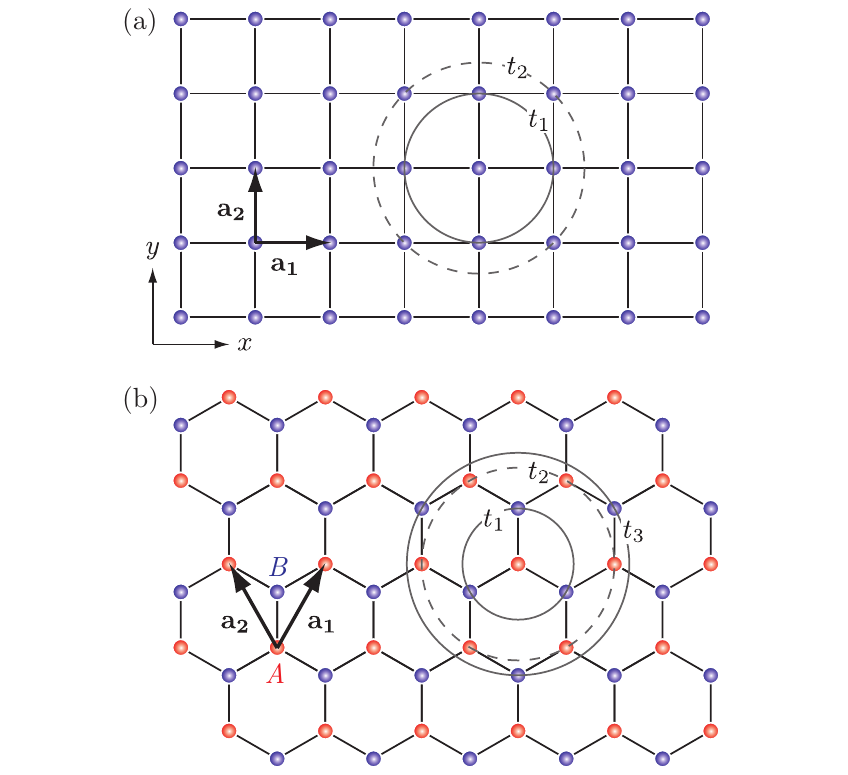}
\caption{\label{fig:Sketch}
(a) Square lattice. (b) Honeycomb lattice with two triangular sublattices $A$ and $B$ (blue and red, respectively). $\mathbf{a_1}$ and $\mathbf{a_2}$ define the unit-cell vector of the Bravais lattices with the lattice constant $a$. $t_1$, $t_2$ and $t_3$ are the N1, N2, and N3 coupling parameters, respectively.}
\end{figure}

\begin{figure}[t]
\centering
\includegraphics[width=8.6cm]{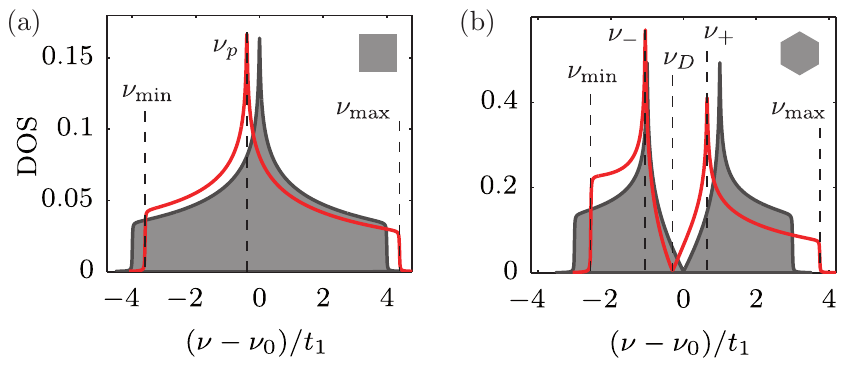}
\caption{\label{fig:DOSanaly} Calculated density of states (DOS) for infinite square (a) and honeycomb (b) lattices. Gray areas: $t_1=-1$ and $t_2 = t_3 = 0$. The spectra are symmetric with respect to $\nu_0$. Red line:  (a) $t_1=-1$, $t_2 = - 0.1$. (b) $t_1=-1$, $t_2 = - 0.1$ and $t_3 = - 0.05$. The positions of the frequencies located by the dashed lines depend on the coupling parameters [see Eqs.~(\ref{eq:Square_numin_numax}, \ref{eq:Square_nup}, \ref{eq:nuD}--\ref{eq:Freq2})].}
\end{figure}

Let us first focus on the square lattice. Since the lattice presents only one site per unit cell [see Fig.~\ref{fig:Sketch}(a)], in the TB approximation, using the Bloch theorem, the dispersion relation can be written as
\begin{equation}
\label{eq:SquareBand}
\nu(\mathbf{k}) - \nu_0 = - \sum_{k} t(\mathbf{R}) e^{i \mathbf{k} \cdot \mathbf{R}}.
\end{equation}
$\mathbf{k} = (k_x,k_y)$ corresponds to the Bloch wave vector, $\mathbf{R}$ is the translation vector of the lattice, and where the on-site resonant frequency $\nu_0$ appears explicitly. $t(\mathbf{R})$ is the coupling between two sites separated by $\mathbf{R}$. If we consider only the coupling terms $t_1$ and $t_2$ between the first and second nearest-neighbors [gray and dashed gray circles in Fig.~\ref{fig:Sketch}(a), respectively], Eq.\eqref{eq:SquareBand} reads\footnote{We use the convention  of Ref.~\onlinecite{CastroNeto2009} regarding the sign of the coupling terms.}
\begin{eqnarray}
\label{eq:SquareNN1}
\nu(\mathbf{k}) - \nu_0 & = & - 2 t_1 \left(\cos{\mathbf{k \cdot a_1}}+ \cos{\mathbf{k \cdot a_2}}\right) \\ \nonumber
& & - 2 t_2 \left[\cos{\mathbf{k \cdot (a_1+a_2)}}+ \cos{\mathbf{k \cdot (a_1-a_2)}} \right]
\end{eqnarray}
where $\mathbf{a_1}$ and $\mathbf{a_2}$ define the primitive cell of the Bravais lattice as depicted in Fig.~\ref{fig:Sketch}(a). The extrema of the energy band correspond to $\mathbf{k \cdot a_1} = \mathbf{k \cdot a_2} = 0$ and $\mathbf{k \cdot a_1} = \mathbf{k \cdot a_2} = \pi$. The width of the band $\Delta \nu$ and its center $\nu_c$ are thus given by
\begin{subequations}
\label{eq:Square_numin_numax}
\begin{eqnarray}
\Delta \nu & = & 8 |t_1| \label{eq:Square_numin} \\
\nu_c & = & \nu_0 - t_2 \label{eq:Square_numax}
\end{eqnarray}
\end{subequations}
The DOS, which is obtained by counting the number of allowed states for each frequency, is non zero between $\nu_{\mathrm{min}} = \nu_c- \Delta \nu /2$ and $\nu_{\mathrm{max}} = \nu_c + \Delta \nu /2$. Moreover, a singularity appears in the DOS at $\nu = \nu_p$. It corresponds to the saddle point in the dispersion relation~\eqref{eq:SquareNN1} which is located at $\mathbf{k \cdot a_1} = 0$ and $\mathbf{k \cdot a_2} = \pm \pi$. We have
\begin{equation}
\label{eq:Square_nup}
\nu_p = \nu_0 + 4 t_2
\end{equation}
The positions of these frequencies depend on the N1 and N2 coupling terms, $t_1$ and $t_2$, respectively. Consequently, as shown in Fig.~\ref{fig:DOSanaly}(a), the shape of the DOS is strongly affected. The spectrum goes from a symmetric distribution (gray area) when only N1 couplings are considered (i.e.,~$t_2=0$) to a non-symmetric shape (red line) when N2 coupling terms are included (i.e.,~$t_2 \neq 0$). The peak shifts and the band extrema $\nu_{\mathrm{min}}$ and $\nu_{\mathrm{max}}$ are modified. Note that the number of states from $\nu_{\mathrm{min}}$ to $\nu_p$ and from $\nu_p$ to $\nu_{\mathrm{max}}$ remains identical.

Let us now focus on the honeycomb arrangement. The situation is different since the lattice is composed of two triangular sublattices $A$ and $B$ [i.e.,~two sites per unit cell, blue and red sites in Fig.~\ref{fig:Sketch}(b)]. The primitive cell of the Bravais lattice is defined by $\mathbf{a_1} = a/2(\sqrt{3},3)$ and $\mathbf{a_2} = a/2(-\sqrt{3},3)$. Starting with an atom on the $A$ lattice, the three N1 (resp. three N3) belong to the $B$ lattice and are located on the smaller (resp. larger) gray circle. The corresponding coupling parameters are $t_1$ and $t_3$ respectively. The six N2 are on the same sublattice and are located on the dashed gray circle.

\begin{table*}[]
\caption{\label{tab:coupling} Coupling parameters obtained, according to Eqs.~\eqref{eq:SquareCoupling1} and~\eqref{eq:Coupling1}, by extracting the values of $\nu_D$, $\nu_p$, $\nu_{\mathrm{min}}$, $\nu_{\mathrm{max}}$, $\nu_-$, and $\nu_+$ from the measured spectra (blue area in Fig.~\ref{fig:DOSexp}). Note that $t_1$, $t_2$, and $t_3$ are negative.}
\begin{ruledtabular}
\begin{tabular}{llllllll}
$d$ (mm)	& $\nu_0$ (GHz)	& $t_1$ (GHz) & $t_2/|t_1|$  & $\nu_0$ (GHz) & $t_1$ (GHz) & $t_2/|t_1|$ & $t_3/|t_1|$\\
\hline
 &\textit{Square lattice} & & & \textit{hc lattice} & & & \\
13			& 6.6603		& -0.0298		& -0.2816		& 6.6535		& -0.0292		& -0.1252		& -0.0324  \\
15			& 6.6549		& -0.0169		& -0.2879		& 6.6563		& -0.0159		& -0.0910		& -0.0709 \\
\end{tabular}
\end{ruledtabular}
\end{table*}

In the Bloch representation, the TB Hamiltonian $\mathcal{H}_{\mathrm{TB}}$ can be written:
\begin{equation}
\label{eq:hamilton}
\mathcal{H}_{\mathrm{TB}} = \begin{pmatrix}
\nu_0  + f_2(\mathbf{k}) 			& f_1(\mathbf{k}) +f_3(\mathbf{k})\\
f_1^*(\mathbf{k}) +f_3^*(\mathbf{k}) 	& \nu_0 + f_2(\mathbf{k})
\end{pmatrix}
\end{equation}
where $f_1$ (resp. $f_2$ and $f_3$) is the first (resp. second and third) nearest-neighbor contribution. For the hc lattice, we can write:
\begin{equation}
\label{eq:NN1}
f_1(\mathbf{k}) = - t_1 \left(1+ e^{i \mathbf{k \cdot a_1}}+ e^{i \mathbf{k \cdot a_2}} \right)
\end{equation}
respectively,
\begin{eqnarray}
\label{eq:NN2}
f_2(\mathbf{k}) = - 2 t_2 \left[\cos{\mathbf{k \cdot a_1}} + \cos{\mathbf{k \cdot a_2}} \right. \\ \nonumber
	\left. + \cos{\mathbf{k \cdot (a_1-a_2)}} \right]
\end{eqnarray}
and
\begin{equation}
\label{eq:NN3}
f_3(\mathbf{k}) = - t_3 \left[ e^{i \mathbf{k \cdot (a_1 + a_2})}+ e^{i \mathbf{k \cdot (a_1 - a_2)}} + e^{i \mathbf{k \cdot (a_2 - a_1)}}\right]
\end{equation}
$\mathbf{k} = (k_x,k_y)$ corresponds to the Bloch wave vector. The energy spectrum is given by:
\begin{equation}
\label{eq:energy}
\nu(\mathbf{k}) - \nu_0 =  f_2(\mathbf{k}) \pm |f_1(\mathbf{k}) + f_3(\mathbf{k})|
\end{equation}
Here, the dispersion relation presents two bands touching at the corners of the Brillouin zone, the so-called Dirac points, for $\mathbf{K \cdot a_1} = \pm 2 \pi/3$ and $\mathbf{K \cdot a_2} = \mp 2 \pi/3$ (so that $f_1$ = $f_3$ = 0). Its energy is therefore
\begin{equation}
\label{eq:nuD}
\nu_D = \nu_0 + 3 t_2
\end{equation}
As depicted in Fig.~\ref{fig:DOSanaly}(b), the DOS vanishes at $\nu=\nu_D$. The extrema of the band energy are obtained when $\mathbf{k \cdot a_1} = \mathbf{k \cdot a_2} = 0$. We have for $\nu_{\mathrm{min}}$ and $\nu_{\mathrm{max}}$
\begin{subequations}
\label{eq:Freq1}
\begin{eqnarray}
\nu_{\mathrm{min}} = \nu_0 - 6 t_2 - 3 |t_1+t_3| \label{eq:numin} \\
\nu_{\mathrm{max}} = \nu_0 - 6 t_2 + 3 |t_1+t_3|  \label{eq:numax}
\end{eqnarray}
\end{subequations}
In addition, the two logarithmic divergences observed in Fig.~\ref{fig:DOSanaly}(b) correspond to the saddle points in the dispersion relation~\eqref{eq:energy} (for $\mathbf{k_M \cdot a_1} = \mathbf{k_M \cdot a_2} = \pi$) and emerge at
\begin{subequations}
\label{eq:Freq2}
\begin{eqnarray}
\nu_- = \nu_0 + 2 t_2 - |t_1 - 3 t_3| \label{eq:nuplus} \\
\nu_+ = \nu_0 + 2 t_2 + |t_1 - 3 t_3| \label{eq:numinus}
\end{eqnarray}
\end{subequations}
Here again, the position of these points depends on the coupling parameters ($t_1$, $t_2$ and $t_3$) and the frequency $\nu_0$. The DOS shape is thus strongly affected as seen in Fig.~\ref{fig:DOSanaly}(b). The two extrema are modified, the vanishing point is shifted, and consequently, the two bands become dissymmetric. We would like to point out that we have neglected the overlap $s$ between nearest-neighbor ($l,  l'$) wave functions: $s=\langle\Psi_l|\Psi_{l'}\rangle \approx 0$. Its effect may be incorporated in a slight change of the $t_i$'s. Therefore,  we consider here that the $t_i$'s are effective coupling parameters.

\subsection{Experimental and analytical DOS}
To experimentally extract the density of states, we average the $g$ function over all positions $\textbf{r}_1$. Indeed, the DOS is directly related to $\langle g(\nu) \rangle _{\textbf{r}_1}$ (see Appendix~\ref{app:gfunction}). As presented in Fig.~\ref{fig:DOSexp}, the spectra have been measured for various lattice constants. Note that, in order to reduce the fluctuations of $\langle g(\nu) \rangle_{\mathbf{r}_1}$ and thus improve the frequency assignment, we use normalized histograms. We choose a bin width of $\Delta \nu_\mathrm{bin} = (1/24) \vert \nu_{\mathrm{max}}-\nu_{\mathrm{min}} \vert$ corresponding to approximately 10 resonances per bin on average. So far, we have not discussed the sign of the couplings. The symmetry of the two-disk system eigenfunctions presented in Sec.~\ref{subsec:TwoDisc} implies $t_1<0$. We observe in Fig.~\ref{fig:DOSexp} that the position of the peak ($\nu_p$) and the vanishing point ($\nu_D$) are shifted towards lower frequencies. In view of Eqs.~\eqref{eq:Square_nup} and \eqref{eq:nuD}, this means that $t_2$ is also negative. Concerning N3 coupling, we show below that the DOS is well fitted when $t_3$ and $t_1$ have the same sign; therefore $t_3 <0$. The common sign for the three nearest-neighbor couplings is consistent with their similar physical origin. \footnote{Note that in the Ref.~\onlinecite{Bellec2013}, one should read
$t=-0.016$ GHz, $t_2/t = 0.091$, $t_3/t= 0.071$.}
\noindent Therefore, by extracting the frequencies of interest from the experimental spectra, we can get, according to Eqs.~\eqref{eq:Square_numin_numax} and \eqref{eq:Square_nup}, the coupling parameters for the square lattice:
\begin{subequations}
\label{eq:SquareCoupling1}
\begin{eqnarray}
\nu_0 & = &\frac{1}{4} (\nu_{\mathrm{min}} + \nu_{\mathrm{max}} + 2 \nu_p)  \label{eq:Square_nu0} \\
|t_1| & = & \frac{1}{8} (\nu_{\mathrm{max}}-\nu_{\mathrm{min}})  \label{eq:Square_t} \\
t_2 & = & \frac{1}{8} (\nu_p - \frac{\nu_{\mathrm{min}} + \nu_{\mathrm{max}}}{2})  \label{eq:Square_t2}
\end{eqnarray}
\end{subequations}
For the hc lattice, according to Eq.~\eqref{eq:Freq1}, if the condition $|t_1| > 3 |t_3|$ is satisfied, we have
\begin{subequations}
\label{eq:Coupling1}
\begin{eqnarray}
\nu_0 & = & \frac{1}{6} (\nu_{\mathrm{min}} + \nu_{\mathrm{max}} + 4 \nu_D)  \label{eq:nu0} \\
|t_1| & = & \frac{1}{8} (\nu_{\mathrm{max}} - \nu_{\mathrm{min}} + \nu_+ - \nu_-)  \label{eq:t} \\
t_2 & = & \frac{1}{9} (\nu_D - \frac{\nu_{\mathrm{min}} + \nu_{\mathrm{max}}}{2})  \label{eq:t2} \\
|t_3| & = & \frac{1}{24} [\nu_{\mathrm{max}} - \nu_{\mathrm{min}} -3(\nu_+ - \nu_-)]  \label{eq:t3}
\end{eqnarray}
\end{subequations}
\begin{figure}[t]
\centering
\includegraphics[width=8.6cm]{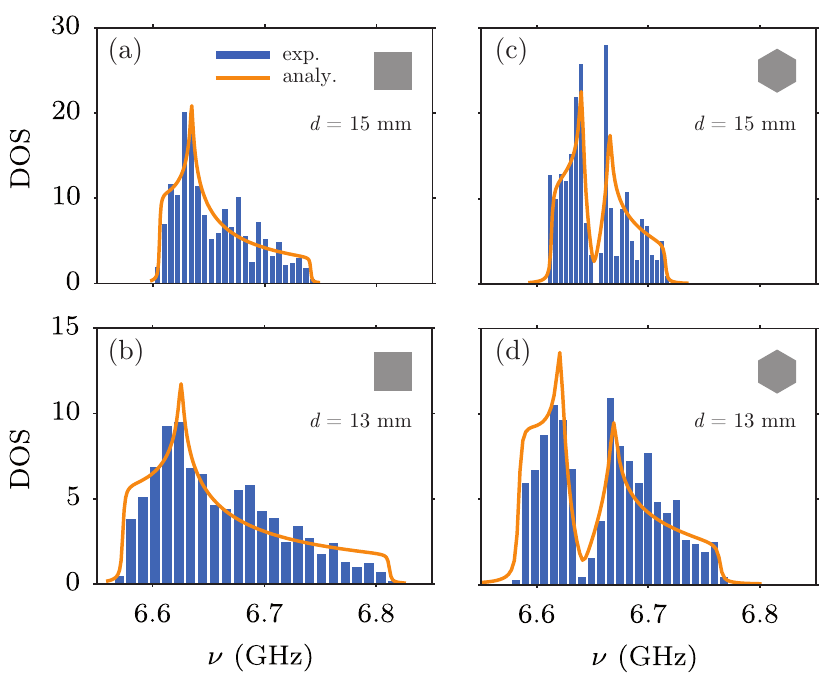}
\caption{\label{fig:DOSexp} DOS for regular square [(a),(b)] and hc [(c),(d)] lattices for various lattice constant $d$. (a), (c) $d=15$\,mm. (b), (d) $d=13$\,mm. Blue area: Normalized histogram with 24 bins per bandwidth (see text for details) of the $g$ function averaged over all the position $\textbf{r}$. Orange line: Analytical solution for a infinite system taking into account the N1, N2, and N3 coupling terms. These parameters are obtained by locating the points of interest as seen in Fig.~\ref{fig:DOSanaly}.}
\end{figure}
The extracted values are reported in Table~\ref{tab:coupling}. The N1 coupling parameters $t_1$ are added in Fig.~\ref{fig:TwoDiscs}(b) for both square (green square) and hc (blue circle) lattice for lattice constant $d$ = 11, 12, 13 and 15\,mm. We observe that, apart from $d$ = 11\,mm, these values are very consistent with the ones obtained with two-disks (gray diamonds) and hexagonal (red circles) systems. Then, the values of the Table~\ref{tab:coupling} are used to numerically calculate the DOS of an \textit{infinite} system in the tight-binding approximation [using Eqs.~\eqref{eq:SquareNN1} and \eqref{eq:energy}]. Note that, to take into account experimental losses, we introduce a Lorentzian broadening in the DOS with a full width at half maximum corresponding to 0.5\% of the bandwidth. The plots are displayed with orange lines in Fig.~\ref{fig:DOSexp}. We observe a good agreement with the experimental data, taking into account that the experimental system is finite (with only $\sim$220 disks). Still, it is possible to observe the dynamic of the spectra which shows the effect of N2 and N3.

Let us focus on the hc lattice [see Figs.~\ref{fig:DOSexp}(c) and \ref{fig:DOSexp}(d)]. For both lattice constants we clearly observe a shift of the Dirac point and a dissymmetric band structure, whereas the number of states remains equivalent in each band. As the first band is narrower, it also becomes more intense, whereas the second band is larger and less intense. The reason is that a large $t_2/t_1$ squeezes considerably the lowest band, and therefore increases the DOS in the lower band (recall that in our system $t_2<0$). As the N2/N1 ratio decreases with $d$, these effects are less significant for the large lattice constant ($d$ = 15\,mm and $t_2/t_1 = 0.09$) than for the smaller one ($d$ = 13\,mm and $t_2/t_1 = 0.12$). Moreover, by increasing $t_2 /t_1$, we observe an increase of the DOS near the lower edge leading to a flattening of the lower band. Actually, one can show that, at $\nu = \nu_\mathrm{min}$,  the DOS increases with $t_2$  as
\begin{equation}
\label{eq:DOSnumin}
\rho(\nu_\mathrm{min}) = {\sqrt{3} \over 2 \pi} {1 \over \sqrt{|t_1|- 6|t_2|}}
\end{equation}
Note that a divergence is expected for $t_2=t_1/6$. Experimentally, since the maximal N2/N1 ratio is $ \simeq 0.125$, we have not been able to reach this critical value. A more detailed analysis of the DOS behavior near the lower edge is presented in Appendix~\ref{app:numerics}.
\section{Conclusion}
\label{sec:Conclusion}
In this paper, we have shown that the propagation of microwaves in an array of dielectric resonators is well described by a tight-binding model, allowing the realization of ``artificial graphene,'' where the microwaves play the role of the electrons in graphene. By changing the distance between resonators, we have  experimentally studied  the role of higher order coupling terms in the frame of the TB regime.  We observe a clear modification of the density of states, with a dissymmetry of the spectrum and a shift of the energy of the Dirac points. Meanwhile, the major characteristics of the DOS -- two bands touching at a (Dirac) point with a vanishing DOS, two logarithmic divergences -- as well as the overall structure of the eigenstates are preserved. This complete characterization of the ``artificial microwave graphene'' may open the way to new experiments in order to easily simulate the fascinating properties of graphene and related systems exhibiting Dirac cones.

\acknowledgments{The authors acknowledge helpful contributions from Chayma Bouazza and Giulia Carra during the early stage of this work.}

\appendix
\section{Details of the $g$-function}
\label{app:gfunction}
{\em Breit-Wigner expression.}---The frequency range used in the experiment being of the order of or less than $200\,$MHz, the coupling to the antenna $\sigma$ may be assumed to be nearly constant.\cite{Barthelemy2005} Thus, the reflection reads ($\mathbf{r}_1$ being the position of the antenna connected to the port 1 of the network analyzer)
\begin{equation}
\label{S11}
S_{11}(\nu)=1-\textrm{i}\sigma G^{+}(\mathbf{r}_1,\mathbf{r}_1;\nu)
\end{equation}
where $G^{+}$ is the regularized Green's function:
\begin{equation}
G^+(\mathbf{r},\mathbf{r};\nu) = \lim_{\Gamma\to 0^+}G(\mathbf{r},\mathbf{r};\nu+\textrm{i}\Gamma)
\end{equation}
The Green's function is the resolvent of the tight-binding Hamiltonian:
\begin{equation}
G(\tilde{\nu})=(\tilde{\nu}\mathbb{I}-\mathcal{H}_{\mathrm{TB}})^{-1}
\end{equation}
where $\tilde{\nu} = \nu + \textrm{i} \Gamma $.\\
By introducing the eigenfunctions $\{\Psi_n(\mathbf{r})\}$ and the eigenvalues $\{\nu_n\}$ of $\mathcal{H}_{\mathrm{TB}}$, expression (\ref{S11}) can be recast in a Breit-Wigner-form, for isolated resonances:
\begin{equation}
\label{BW}
S_{11}(\nu)=1-\textrm{i}\sigma\sum_n\frac{\vert\Psi_n(\mathbf{r}_1)\vert^2}{\nu-\nu_n+\textrm{i}\Gamma}
\end{equation}
One can legitimately assume that the  width $\Gamma$ corresponds to \emph{homogeneous} damping. Indeed, since the damping is essentially due to the Ohmic losses in the bottom and top metallic plates sandwiching the dielectric resonators, a constant and uniform decay rate for all the eigenmodes ($\Gamma_n\equiv\Gamma$) is expected.\cite{Barthelemy2005,Barthelemy2005a,*Savin2006} The local density of states is given by:
\begin{equation}
\rho(\mathbf{r},\nu)=-\frac{1}{\pi}\mathcal{I}m\,G^+(\mathbf{r},\mathbf{r};\nu)=\sum_n\vert\Psi_n(\mathbf{r})\vert^2\delta(\nu-\nu_n)
\end{equation}

{\em Isolated resonance.}---For an isolated resonance, the sum in (\ref{BW}) contains only one term
\begin{equation}
S_{11}(\nu)=1-\textrm{i}\sigma\frac{\vert\Psi_0(\mathbf{r})\vert^2}{\nu-\nu_0+\textrm{i}\Gamma}
\end{equation}
Thus, the amplitude of the reflected signal -- the network analyzer does not fix an absolute phase reference -- can be related to the intensity of the wave function [neglecting the $(\sigma/\Gamma)^2$ term]:
\begin{equation}
1-\vert S_{11}(\nu)\vert^2 \simeq \frac{2\sigma\Gamma}{\left(\nu-\nu_0\right)^2+\Gamma^2}\,\vert\Psi_0(\mathbf{r}_1)\vert^2
\end{equation}
Close to the eigenfrequency $\nu_0$, one has $1-\vert S_{11}(\nu_0)\vert^2\simeq (2\sigma/\Gamma)\,\vert\Psi_0(\mathbf{r}_1)\vert^2$

{\em $g$ function.}---One defines the ``$g$ function'' by
\begin{equation}
g(\mathbf{r}_1,\nu) = \frac{\vert S_{11}(\nu)\vert^2}{\langle \vert S_{11}\vert^2 \rangle_{\nu}} \varphi_{11}'(\nu)
\end{equation}
where $\langle\ldots\rangle_\nu$ indicates an averaging over the whole range of the frequency spectra, $\varphi_{11}$ is the phase of the reflected signal: $\varphi_{11}=\mathrm{Arg}(S_{11})$, and where $\varphi_{11}'$ denotes its derivative with respect to the frequency. To avoid non physical singularities in the derivative due to the modulo-$\pi$ occurring in the $\arctan$ function, we use the following expression of $\varphi_{11}'$:
\begin{equation}
\varphi_{11}' = \frac{\mathcal{I}m(S'_{11})\mathcal{R}e(S_{11}) - \mathcal{I}m(S_{11})\mathcal{R}e(S'_{11}) }{|S_{11}|^2}
\end{equation}
In the regime of non overlapping resonances, this quantity takes non zero values only in the vicinity of the eigenfrequencies. Close to a given eigenfrequency, the real and imaginary part of $S_{11}^{\prime}$
\begin{equation}
S_{11}^{\prime}(\nu)=\textrm{i}\,\sigma\sum_n\frac{\vert\Psi_n(\mathbf{r}_1)\vert^2}{(\nu-\nu_n+\textrm{i}\Gamma)^2}
\end{equation}
exhibit the following dominant behaviors ($\nu\simeq\nu_n$):
\begin{equation}
\mathcal{R}e\, S_{11}^{\prime}(\nu_n)\sim 0\qquad\mathcal{I}m\, S_{11}^{\prime}(\nu_n)\simeq -\sigma\frac{\vert\Psi_n(\mathbf{r}_1)\vert^2}{\Gamma^2}
\end{equation}
It follows that
\begin{equation}
\varphi_{11}'(\nu_n)\simeq  -\frac{\sigma}{|S_{11}|^2}\frac{\vert\Psi_n(\mathbf{r}_1)\vert^2}{\Gamma^2}
\end{equation}
and
\begin{equation}
g(\mathbf{r}_1,\nu) = -\frac{\sigma}{\Gamma\,\langle \vert S_{11}\vert^2 \rangle_{\nu}}\sum_n\vert\Psi_n(\mathbf{r}_1)\vert^2\frac{\delta_{\nu,\nu_n}}{\Gamma}
\end{equation}
Having in mind that $\Gamma$, the resonance width, gives the frequency resolution, the quantity $\delta_{\nu,\nu_n}/\Gamma$ can be viewed as a discrete version of the  delta function $\delta(\nu-\nu_n)$. Thus, we obtain through the $g$-function an approximated evaluation of the density of states:
\begin{eqnarray}
g(\mathbf{r}_1,\nu) &=& -\frac{\sigma}{\Gamma\,\langle \vert S_{11}\vert^2 \rangle_{\nu}}\sum_n\vert\Psi_n(\mathbf{r}_1)\vert^2\delta(\nu-\nu_n)\\
&=&-\frac{\sigma}{\Gamma\,\langle \vert S_{11}\vert^2 \rangle_{\nu}}\rho(\mathbf{r}_1,\nu)
\end{eqnarray}
We get $\vert \Psi_n(\mathbf{r}_1)\vert^2$ by taking $\displaystyle{\max_{\nu \approx \nu_0}(g(\mathbf{r}_1,\nu))}$ for each position $\mathbf{r}_1$. Note that $\displaystyle{-\frac{\sigma}{\Gamma\,\langle \vert S_{11}\vert^2 \rangle_{\nu}}}$ renormalizes the effects of the baseline coming from the other resonances (when $i \neq n$) and from experimental artifacts.

\section{Van Hove singularity at the lower band edge}
\label{app:numerics}
\begin{figure}[t]
\centering
\includegraphics[width=8.6cm]{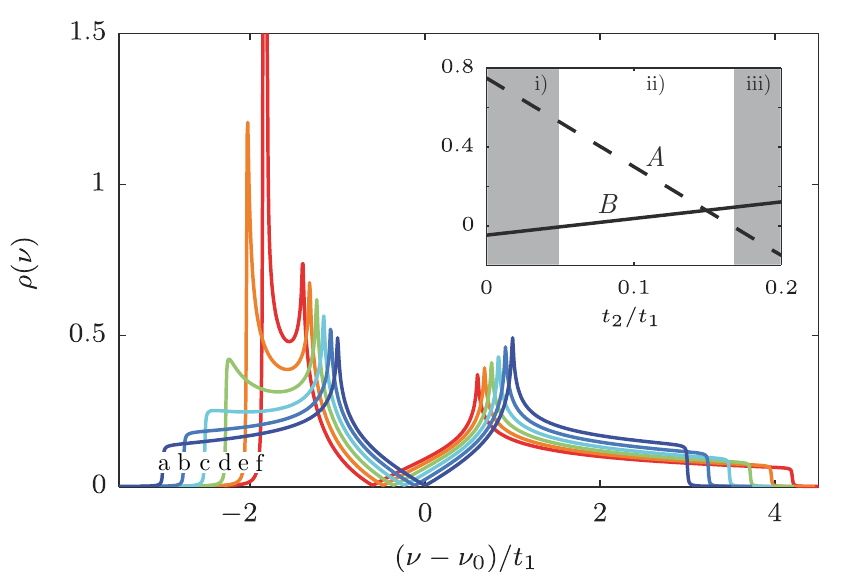}
\caption{\label{fig:DOSnoB} Density of states for $t_2/t_1=0.0$ (a), 0.04 (b), 0.08 (c), 0.12 (d), 0.16 (e), and 0.2 (f) with $t_1 = -1$ and $t_3=0$. The DOS is calculated numerically with minimal broadening. Inset: $A$ and $B$ parameters versus $t_2/t_1$.  }
\end{figure}
Figure~\ref{fig:DOSnoB} presents the DOS for various $t_2/t_1$ ranging from 0 to 0.2, where $t_1 = - 1$ and $t_3 = 0$ for simplicity (minimal broadening here). For large $t_2$, a divergence of the DOS appears at the lower edge of the spectrum. The reason is the following: Near the lower edge of the spectrum, that is around  $k = 0$, we have
\begin{equation}
\nu(\mathbf{k}) = \nu_{\mathrm{min}} + A \mathbf{k}^2  + B \mathbf{k}^4
\end{equation}
with $A =  3 |t_1|/4 - 9 |t_2| /2 $ and $B =  -3 |t_1| /64  + 27 |t_2| / 32$. The density of states per unit cell reads
\begin{equation}
\rho(\nu) = \frac{3 \sqrt{3}}{8 \pi} \frac{1}{\sqrt{A^2 + 4 B (\nu - \nu_\mathrm{min})}}
\end{equation}
At $\nu=\nu_\mathrm{min}$,  we find the expression~(\ref{eq:DOSnumin}). As $|t_2|/|t_1|$ increases, $A$ and $B$ evolves and several behaviors can be identified (see inset in Fig.~\ref{fig:DOSnoB}).

(i) For $|t_2|/|t_1| < 1/18$, we have $A>0$ and $B<0$. Since $B \ll A$, the DOS is almost constant near the edge (as expected for a quadratic dispersion relation in 2D), with a linear correction.
\begin{equation}
\rho(\nu) = \frac{3 \sqrt{3}}{8 \pi} \left(1-\frac{2B}{A^2}(\nu-\nu_\mathrm{min}) \right)
\end{equation}
The linear term becomes negative for $1/18 <| t_2|/|t_1| < 1/6$ (i.e., when $A > 0$ and $B > 0$). Such evolutions are observed in the Figs.~\ref{fig:DOSnoB}(a)-\ref{fig:DOSnoB}(d).

(ii) At the critical point, for $|t_2|/|t_1| = 1/6$ we have $A=0$. The DOS exhibits the following square root behavior:
\begin{equation}
\rho(\nu) = \frac{3 \sqrt{3}}{8 \pi} \frac{1}{\sqrt{4 B (\nu - \nu_\mathrm{min})}}
\end{equation}
At $\nu=\nu_\mathrm{min}$, as observed in Fig.~\ref{fig:DOSnoB}(e), the DOS diverges.

(iii) For $|t_2|/|t_1| > 1/6$, the square root behavior still remains [see Fig.~\ref{fig:DOSnoB}(f)]:
\begin{equation}
\rho(\nu) = \frac{3 \sqrt{3}}{8 \pi} \frac{1}{\sqrt{4 B (\nu - \nu_1)}}
\end{equation}
with $\nu_1 =  \nu_\mathrm{min} - A^2 / 4B$. Similarly, one can show that, in the case of the square lattice, for $t_2< t_1/2$, the DOS at $\nu = \nu_\mathrm{min}$ increases as follows:
\begin{equation}
\rho(\nu_\mathrm{min})  = {3 \sqrt{3} \over 8 \pi \sqrt{|t_1|- 2 |t_2|}}
\end{equation}
Therefore, in view of Eq.~(\ref{eq:DOSnumin}), it increases much slower than for the hc lattice as seen by comparing Fig.~\ref{fig:DOSexp}(a), \ref{fig:DOSexp}(b) and Figs.~\ref{fig:DOSexp}(c), \ref{fig:DOSexp}(d).

%

\end{document}